\documentstyle[12pt]{article}

\begin{document}

{\footnotesize {hep-th/9511187 {\hfill} USC-95/HEP-B3}} \bigskip

\begin{center}
{\normalsize {\bf SOLUTION OF THE SL(2,R) STRING}}

{\normalsize {\bf IN CURVED SPACETIME}}\footnote{%
Based on lectures delivered at the Strings '95 conference, USC, March 1995,
and at the Strings, Gravity and Physics at the Planck Scale conference,
Erice, August 1995.}

\baselineskip=22pt
\end{center}

\centerline{\footnotesize ITZHAK BARS} \baselineskip=13pt

\centerline{\footnotesize {\it Department of Physics and Astronomy,
University of Southern California}} \baselineskip=12pt \centerline{%
{\footnotesize {\it Los Angeles, CA 90089-0484, USA}}}

\centerline{\footnotesize E-mail: bars@physics.usc.edu}

\vspace*{0.9cm} {\ \centering{%
\begin{minipage}{12.2truecm}\footnotesize\baselineskip=12pt\noindent
\centerline{\footnotesize ABSTRACT}\vspace*{0.3cm}
\parindent=0pt \
The SL(2,R) WZW model, one of the simplest models for strings propagating
in curved space time, was believed to be non-unitary in the algebraic
treatment involving affine current algebra. It is shown that this was an
error that resulted from neglecting a zero mode that must be included to
describe the correct physics of non-compact WZW models. In the presence
of the zero mode the mass-shell condition is altered and unitarity is restored.
The correct currents, including the zero mode, have logarithmic cuts
on the worldsheet. This has physical consequences for the spectrum
because a combination of zero modes must be quantized in order to impose
periodic boundary conditions on mass shell in the physical sector of the
theory.
To arrive at these results and to solve the model completely, the SL(2,R) WZW
model is quantized in a new free field formalism that differs from previous
ones
in that the fields and the currents are Hermitean, there are cuts, and there is
a
new term that could be present more generally, but is excluded in the WZW
model.
\end{minipage}}}


\baselineskip=15pt \setcounter{footnote}{0} \renewcommand{\thefootnote}{%
\alph{footnote}}


\section{Introduction}

One of the main reasons to study string theory in curved spacetime is to
develop the appropriate methods to investigate physical phenomena in the
presence of quantum gravity in a mathematically consistent theory. Quantum
gravity is important during the early universe and this must have an impact
on the symmetries and matter content (gauge bosons, families of quarks and
leptons) observed at accelerator energies. In addition, in order to develop
an understanding of gravitational singularities such as black holes or the
big bang, including quantum aspects of gravity, string theory in curved
spacetime must be investigated.

A string propagating in curved{\normalsize \ }spacetime with one time and $%
(d-1)$ space coordinates is described by a string action that has the
following form in the conformal gauge
\begin{equation}
S=\int d\tau d\sigma \,[\partial _{+}X^\mu \partial _{-}X^\nu \,G_{\mu \nu
}(X)+\cdots ]  \label{action}
\end{equation}
where $G_{\mu \nu }(X)$ is a background metric in $d$-dimensions with
signature $(-1,1,1,\cdots ).$ The terms in the action denoted by $\cdots $
may contain additional background fields such as an antisymmetric tensor $%
B_{\mu \nu }(X)$ a dilaton $\Phi (X)$ etc.. Conformal invariance must be
imposed on these background fields at the quantum level, otherwise
reparametrization invariance (Virasoro constraints) cannot be used to remove
ghosts from the theory.

In one time plus $(d-1)$ space dimensions many models that are exactly
conformally invariant at the quantum level have been constructed by now \cite
{BN}\cite{ibcurved}\cite{tseytlinrev}. A first example of string propagation
in curved spacetime was the SL(2,R) WZW model that potentially could be
solved through algebraic methods. However, one immediately came across an
unexpected inconsistency problem involving the unitarity of the theory \cite
{BN}\cite{O'R}. It was noticed a long time ago that the Virasoro constraints
were insufficient to remove all the negative norm states encountered in the
affine current algebra treatment of the model. There seemed to be more
negative norm states than those introduced by the time-like string
coordinates. This problem persited with all other non-compact affine current
algebra coset models, including the much studied SL(2,R)/R two dimensional
black hole \cite{BN}\cite{WIT}. This observation casted doubt on the
physical validity of the models and prevented the development of physical
ideas based on the algebraic properties of these (in principle) completely
solvable models.

In this talk, and in a related paper \cite{ibsl2wzw}, I explain the solution
of the unitarity problem. I show that the WZW model has more degrees of
freedom than those described by the affine current algebra. The extra
degrees of freedom are zero modes (in particular, in non-compact
directions). In the old treatment one inadvertently set the zero mode to
zero values, and hence missed important properties of the model. One of the
important effects of the non-trivial zero mode is that the mass shell
condition is altered. In the absence of the zero mode the string cannot be
put correctly on mass shell and this begins to explain why the non-unitary
states emerged.

To actually show that the model has no ghosts a second step is needed. In
particular one must argue that the discrete series representation of SL(2,R)
does not appear at the base, since in the purely algebraic approach
(including the new zero mode) there would still be negative norm states in
this representation, according to the old approach. To show that the WZW
model excludes the discrete series, I formulate the quantum theory in terms
of a suitable parametrization of SL(2,R) that corresponds to free fields,
and then show that only the unitary principal series is allowed. In this
formulation the spectrum is completely solved and the Virasoro constraints
implemented. The no ghost theorem is then shown to be valid.

In the presence of non trivial values of the zero mode (which is needed for
the correct physics) the conserved currents of the theory have logarithmic
singularities on the world sheet. {\it A priori} one may think that this
would imply that the closed string boundary conditions are not satisfied.
However, this is not the case, because periodicity is required only on mass
shell. In the presence of the new zero modes the Hilbert space is larger
than before. Imposing periodicity gives quantized values for a combination
of the zero modes, thus satisfying the correct closed (or open) string
boundary conditions for the physical on-mass-shell states.

To arrive at these results, and also to solve the model completely, a new
free field formalism is introduced as mentioned above. The structure of the
currents in terms of the free fields is reminiscent of the one found by
Wakimoto \cite{wakim} and Gerasimov et. al. \cite{gerasim}, but it differs
from theirs in three important aspects: (1) The free fields as well as the
currents are Hermitean; this is important for the discussion of unitarity.
(2) The currents have logaritmic cuts that are associated naturally with the
zero modes of the free fields; this affects the spectrum and monodromy. (3)
A new term, that can be present in the free field formulation to reproduce
the effects of the most general SL(2,R) currents, is introduced in order to
obtain all unitary representations at the base. The source of all the
unwanted ghosts is traced to the new term. But, for the specific model at
hand, i.e. the WZW model, the extra term is shown to be absent, thus
elucidating the mechanism by which the model becomes unitary.

\section{The unitarity problem}

The SL(2,R) WZW model has a timelike coordinate that introduces negative
norm oscillators. These create negative norm states, but this is not the
source of the problem. On the basis of na\"{\i}ve counting one may hope that
the Virasoro constraints will remove these ghosts from the theory (this
expectation is born out in our final result). A similar situation occurs
also in the flat theory. As is well known, in the flat case one can indeed
prove the no ghost theorem \cite{noghost} which implies that the theory is
unitary. However, in the case of curved spacetime current algebra models,
one finds that even after imposing the Virasoro constraints there remains
negative norm states that render the theory non-unitary. This has been the
main stumbling block that discouraged the application of these ideas to
model building for the past five years.

Negative norm states that satisfy the Virasoro constraints can be displayed
explicitly. An example is \cite{BN}
\begin{eqnarray}
\begin{array}{l}
|\phi ,l>=\left( J_{-1}^1-iJ_{-1}^2\right) ^l\,|j,m=j+1>, \\
\tilde{L}_n|\phi ,l>=0,\quad \quad n\geq 1, \\
<\phi ,l|\phi ,l>=N_{j(l)}\,\left( l!\right) \prod_{r=0}^{l-1}\left(
k-2j(l)-2+r\right) .
\end{array}
\label{negnorm}
\end{eqnarray}
where $N_{j(l)}=<j,m=j+1\,|j,m=j+1>$ is the norm of the state at the base.
The base is in the discrete series representation of SL(2,R), with $m\geq
(j+1).$ It is required to be in the discrete series\footnote{%
When $j(j+1)>0$ only the discrete series can occur among the unitary
representations of SL(2,R). By contrast, the principal series occurs only
when $j(j+1)<-1/4.$} by the mass-shell condition $L_0=1,$ which gives
\begin{equation}
-\frac{j(j+1)}{k-2}+l=1.  \label{m-shell}
\end{equation}
Evidently, for sufficiently large values of the excitation number $l$ the
norm switches between positive and negative values. Hence, despite the
Virasoro constraints this model is not unitary and cannot describe a
physical string.

\section{Solution of the unitarity problem}

Until now a solution to this problem, and the related SL(2,R)/R black hole
problem, has not been found despite many attempts \cite{ibstonybrook}\cite
{attempts}. Suggestions included: (1) Restrict (artificially) $j(l)+1<k/2$
so that the norm never becomes negative; (2) Allow large values of $j(l)$ as
needed by the excited level $l,\,$ but also permit the base state to have
negative norm $N_{j(l)}$ in such a way as to make the norm of the excited
state $<\phi ,l|\phi ,l>$ positive; (3) Hope that modular invariants will
fix the problem. All of these suggestions are rejected \cite{ibsl2wzw}.

The resolution of the problem lies in understanding that wrong assumptions
have been made about the algebraic structure of the WZW model. In
particular, the assumption that the SL(2,R) WZW model is described by affine
SL(2,R) is not entirely correct. There is an additional zero mode that is
present in the local conserved SL(2,R) currents of the WZW model, whose
presence is crucial for the resolution of the unitarity issues. This zero
mode is missed by the assumption of affine currents that are written in the
form of a Laurent series $\tilde{J}^a(z)=\sum J_n^a\,z^{-n-1}$. When the
additional zero mode is included, the true currents $J^a(z)$ have a
logarithmic term $\ln z$ in addition to the usual powers $z^n.$ Hence,
manipulations such as (\ref{negnorm},\ref{m-shell}) based on the old affine
currents $\tilde{J}^a(z)$ do not fully reflect the correct theory, and this
is why we find inadmissible unphysical results.

In general one can include such $\ln z$ parts, but still have the correct
local commutation rules or operator products with only poles. To see this,
let $\tilde{J}^a(z)$ be the usual Laurent series (with the usual operator
products) that have modes $J_n^a$ as above, and in addition introduce a new
zero mode $\alpha _0^{-}$ that commutes with all the other modes $J_n^a.$
The new currents are
\begin{eqnarray}
\begin{array}{l}
J^0(z)+J^1(z)=\left[ \tilde{J}^0(z)+\tilde{J}^1(z)\right] \\
J^0(z)-J^1(z)=\left[ \tilde{J}^0(z)-\tilde{J}^1(z)\right] -2i\alpha
_0^{-}\ln z\,\,\tilde{J}^2(z) \\
\quad \quad -\frac kz\alpha _0^{-}+\left( -i\alpha _0^{-}\ln z\right)
^2\left[ \tilde{J}^0(z)+\tilde{J}^1(z)\right] \\
J^2(z)=\tilde{J}^2(z)-i\alpha _0^{-}\ln z\,\left[ \tilde{J}^0(z)+\tilde{J}%
^1(z)\right]
\end{array}
\label{newold}
\end{eqnarray}
It can be shown that they have the usual correct operator products, with
only poles, for any value of the zero mode $\alpha _0^{-}$ \cite{ibsl2wzw}$.$

The $\ln z$ terms arise naturally in the canonical formulation of the WZW
model. There are left/right moving string coordinates $X_{L,R}$ that
parametrize the group element $g(X_L(z),X_R(\bar{z})).$ As usual, string
coordinates have a ``momentum'' zero mode $p\ln z$. The currents in this
model depend both on $X_{L,R}$ as well as on their derivatives. Therefore,
the currents are expected to include $\ln z$ pieces proportional to the zero
modes. This is unlike the flat theory, where the currents depend only on the
derivatives of the $X_{L,R}.$ If one sets the $p\ln z$ parts equal to zero,
as is inadverdently done by assuming affine currents in the form of Laurent
series, one forces the string to lie in a sector of fixed zero mode $p=0.$
This may be harmless in compact directions, but it is fatal in non-compact
directions. For example, for the flat string, if one requires the lightcone
momentum $p^{-}=0,$ then the mass-shell condition
\[
p^{+}p^{-}-p_i^2=l-1
\]
cannot be satisfied in that sector. In the SL(2,R) string exactly this
situation arises when $\alpha _0^{-}=0$. It turns out that, in SL(2,R), in
the absence of the zero mode, which is analogous to $p^{-},$ it is still
possible to satisfy a mass shell condition, but only in the discrete series
representation. As seen above the discrete series gives rise to ghosts. On
the other hand, when the zero mode $\alpha _0^{-}\neq 0$ is included (and
hence $\ln z$ is present in the currents), the mass shell condition is
satisfied in the principal series representation, which is free of ghosts at
the excited levels.

\section{SL(2,R) Currents and free fields}

Consider the free fields $X^{-}(z),\,P^{+}(z),$ $\,S(z),\,T^{\prime }(z)$.
They have na\"{\i}ve dimensions 0,1,1,2 respectively and they are defined as
follows

\begin{eqnarray}
X^{-}(z) &=&q^{-}-i\alpha _0^{-}\,\ln z+i\sum_{n\neq 0}\frac 1n\alpha
_n^{-}\,z^{-n},\,\quad \left( \alpha _n^{-}\right) ^{\dagger }=\alpha
_{-n}^{-}  \nonumber \\
P^{+}(z) &=&\sum_{n=-\infty }^\infty \alpha _n^{+}\,z^{-n-1},\,\quad \left(
\alpha _n^{+}\right) ^{\dagger }=\alpha _{-n}^{+}  \nonumber \\
S(z) &=&\sum_{n=-\infty }^\infty s_n\,z^{-n-1},\quad \left( s_n\right)
^{\dagger }=s_{-n}  \label{fields} \\
T^{\prime }(z) &=&\sum_{n=-\infty }^\infty L_n^{\prime }\,z^{-n-2},\quad
\left( L_n^{\prime }\right) ^{\dagger }=L_{-n}^{\prime }  \nonumber
\end{eqnarray}
These fields are Hermitean. The currents are
\begin{eqnarray}
J_0(z)+J_1(z) &=&\,P^{+}(z)  \nonumber \\
J_0(z)-J_1(z) &=&\,:X^{-}(z)\,P^{+}(z)\,X^{-}(z):+2S(z)\,X^{-}(z)  \nonumber
\\
&&-\,\,ik\partial _zX^{-}(z)-\frac{(k-2)T^{\prime }(z)}{P^{+}(z)}
\label{currents} \\
J_2(z) &=&\,:X^{-}(z)\,P^{+}(z):+S(z)  \nonumber
\end{eqnarray}
The currents have a Wakimoto type structure, but with the exception that
these currents are Hermitean. Furthermore, they contain two other new
features: (i) the $\ln z$ parts that have the same structure as (\ref{newold}%
), and (ii) the new terms $T^{\prime }/P^{+}.$ It can be shown that the
operator products of free fields produce the correct operator products of
the currents \cite{ibsl2wzw}.

The free field commutation rules are
\begin{equation}
\begin{array}{l}
\lbrack q^{-},\alpha _0^{+}]=i\,, \\
\left[ \alpha _n^{-}\,,\alpha _m^{+}\right] =n\,\delta _{n+m,0}\,\,\,, \\
\left[ s_n\,,s_m\right] =\left( \frac k2-1\right) \,n\,\delta _{n+m,0}\,\,\,,
\\
\left[ L_n^{\prime }\,,L_m^{\prime }\right] =(n-m)L_{n+m}^{\prime }+0
\end{array}
\label{commutators2}
\end{equation}
while all other commutators are zero. The $\alpha _n^{\pm }$ oscillators may
be rewritten in terms of light-cone type combinations of one time-like $%
\alpha _n^0$ and one space-like $\alpha _n^1$ oscillator, i.e. $\alpha
_n^{\pm }=(\alpha _n^1\pm \alpha _n^0)/\sqrt{2}.$ This shows one source of
negative norms, but by naive counting, they are expected to be removed by
the Virasoro constraints. In fact, the usual proof of the no-ghost theorem
can be extended to show that these ghosts are indeed removed \cite{ibsl2wzw}%
. The zero modes $q^{-},p^{\pm }$ are interpreted as light-cone type
canonical variables $q=x^{-},\,p=p^{+}.$ We have not introduced a canonical
variable corresponding to $x^{+},$ hence $\alpha _0^{-}=p^{-}$ commutes with
all$\,$ the operators and acts like a constant. Similarly the zero mode $s_0$
also acts like a constant. $\alpha _0^{\pm },s_0,L_0^{\prime }$ are
simultaneously diagonalized in the Hilbert space, and they label the base.

The $L_n^{\prime }$ operators act like Virasoro operators with zero central
charge. It is always possible to construct such an operator in terms of free
fields, the simplest being a (negative norm) free boson with a background
charge. Another example is any critical conformal field theory including the
conformal ghosts with central charge $c=-26$. Both of these examples have
ghosts and this is true more generally. Indeed, there is no construction of
a zero central charge Virasoro operator that does not contain negative norm
states in its Hilbert space. Hence, in the presence of $L_n^{\prime }$ there
is an additional source of negative norms that will not be possible to be
removed by the Virasoro constraints. For example a base state with the
property $L_0^{\prime }|h^{\prime }>=h^{\prime }|h^{\prime }>$ produces a
negative norm state $|\phi >=L_{-1}^{\prime }|h^{\prime }>,$
$<\phi |\phi >=2h^{\prime }<h^{\prime }|h^{\prime }>,$ if $h^{\prime }<0.$
As we will see below only when $L^{\prime }$ is present and $h^{\prime }<0,$
the discrete series is possible. Hence the unwanted negative norm states and
the discrete series go hand-in-hand. We will show that $L^{\prime }$ is
absent in the WZW model, hence the WZW model has no additional sources of
negative norms.

\section{Stress tensor and free fields}

The energy momentum tensor is obtained from the normal ordered product of
the currents
\begin{equation}
T(z)=\frac 1{k-2}:\left( -\left( J_0(z)\right) ^2+\left( J_1(z)\right)
^2+\left( J_2(z)\right) ^2\right) :  \label{stresst}
\end{equation}
The result of the computation gives
\begin{equation}
T(z)=\,:P^{+}i\partial _zX^{-}:+T_S(z)+T^{\prime }(z)\quad ,
\label{stresstensor}
\end{equation}
If the computation is repeated with the $\tilde{J}(z)$ currents the only
difference is dropping the $\alpha _0^{-}$ term contained in
\begin{equation}
i\partial _zX^{-}=\sum_{n=-\infty }^\infty \alpha _n^{-}z^{-n-1}.
\end{equation}
In (\ref{stresstensor}) $T_S$ is a Hermitean stress tensor
\begin{equation}
T_S(z)=\frac 1{k-2}\left[ :\left( S(z)\right) ^2:-\,\,\frac iz\partial
_z\left( zS(z)\right) +\frac 1{4z^2}\right]  \label{ts}
\end{equation}
The structure $\frac iz\partial _z\left( zS(z)\right) $ differs from the
usual one $i\partial S,$ and thus is Hermitean. The operator products of $%
T_S(z)$ are

\begin{equation}
T_S(z)\times T_S(w)=\frac{c_s/2}{\left( z-w\right) ^2}+\frac{2T_S(w)}{\left(
z-w\right) ^2}+\frac{\partial _wT_S(w)}{z-w}+\cdots
\end{equation}
with the central charge
\begin{equation}
c_s=1+\frac 6{k-2}.
\end{equation}
Note that the term $P^{+}i\partial _zX^{-}$ is identical to the energy
momentum tensor of flat light-cone coordinates constructed from the
oscillators $\alpha _n^{\pm }$. Therefore, that part is mathematically
equivalent to a $c=2$ stress tensor constructed from one time and one space
coordinate in flat spacetime. Then the total central charge is
\begin{eqnarray}
&&
\begin{array}{ll}
c & =2+c_s+c^{\prime } \\
& =2+\left( 1+\frac 6{k-2}\right) +0 \\
& =\frac{3k}{k-2},
\end{array}
\end{eqnarray}
which is the right central charge for the $SL(2,R)$ WZW model. Finally, as a
further consistency check, by using only the operator products of the
elementary fields, one finds that $T(z)$ has the correct operator products
with the currents.

The zero mode of the stress tensor takes the form $L_0=L_0^{\pm
}+L_0^S+L_0^{\prime },$ where each piece has the eigenvalues
\begin{eqnarray}
&&
\begin{array}{ll}
L_0^{\pm }= & p^{+}p^{-}+l_{\pm }, \\
L_0^S= & (s_0^2+1/4)/(k-2)+l_s, \\
L_0^{\prime }= & h^{\prime }+l^{\prime }.
\end{array}
\end{eqnarray}
where $l_{\pm },l_s,l^{\prime }$ are positive integers and $h^{\prime }$ is
the eigenvalue of $L_0^{\prime }$ at the base (whose possible values depend
on the model for $T^{\prime })$. The mass shell condition $L_0=a$ is
\begin{equation}
p^{+}p^{-}+(s_0^2+1/4)/(k-2)+h^{\prime }+\ {\rm integer}=a
\label{balancee}
\end{equation}
where $a\leq 1.$ The term $p^{+}p^{-}$ is crucial since it takes negative
values, as seen below.

To identify the value of the Casimir operator $j(j+1)$ associated with the
affine currents $\tilde{J}$ in (\ref{newold}) we set $p^{-}=0,$ and compare
the eigenvalue of the resulting $L_0$ to the standard formula. We then see
that the Casimir of the old currents ($\tilde{J}_0)^2=-j(j+1)$ takes the
value
\begin{equation}
j(j+1)=-(s_0^2+1/4)-h^{\prime }(k-2).
\end{equation}
Therefore, if the $T^{\prime }$ piece is absent in the construction ($%
h^{\prime }=0),$ then $j=-1/2\pm is_0$ is only in the principal series. The
supplementary series could occur for $-1/4<j(j+1)<0$ and the discrete series
occurs for $-1/4<j(j+1).\,$ We see that the field $T^{\prime }$ with a
positive $h^{\prime }$ contributes only to the principal series and with a
negative $h^{\prime }$ it leads to the other representations as well. This
construction may find various applications in the future. We will see below
that $T^{\prime }$ is absent in the SL(2,R) WZW model, hence only the
special case of our construction ($T^{\prime }=0)\,$ finds an application in
the WZW model. Then, for excited string states, since the integer in (\ref
{balancee}) is positive it would not be possible to satisfy the mass shell
condition in the absence of the $p^{-}.$ So, the logarithmic structure plays
a role.

\section{The SL(2,R) WZW model}

We now relate the algebraic structures above to the WZW model for SL(2,R).

The quantum theory for any WZW model at the critical point is conveniently
formulated in terms of the left and right moving currents after writing the
group element $g(\tau ,\sigma )=g_L(\tau +\sigma )\,g_R^{-1}(\tau -\sigma )$
\begin{equation}
J_L(z)=ik\partial _zg_Lg_L^{-1},\quad J_R(\bar{z})=ik\partial _{\bar{z}%
}g_Rg_R^{-1},  \label{clcur}
\end{equation}
where $z=e^{i(\tau +\sigma )},\,\bar{z}=e^{i(\tau -\sigma )}.\,\,$The
quantum rules are most conveniently given in terms of operator products
among the currents and the group elements.
\begin{eqnarray}
J_{L,R}^i(z)\,J_{L,R}^j(w) &\rightarrow &\frac{k/2}{\left( z-w\right) ^2}%
+i\epsilon ^{ijl}\,\eta _{lk}\frac{J_{L,R}^k\left( w\right) }{z-w}+\cdots
\label{quantumrules} \\
J_{L,R}^i(z)\,g_{L,R}(w) &\rightarrow &\frac{-t^i}{z-w}g_{L,R}(w)+\cdots
\nonumber
\end{eqnarray}
The $t_i$ is a basis for the SL(2,R) Lie algebra which is given in terms of
Pauli matrices $t_0=\sigma _2/2,\,\,t_1=i\sigma _1/2,\,\,t_2=-i\sigma _3/2.$
They satisfy $\eta _{ij}=-2tr(t_it_j)=diag(-1,1,1).$

Any group element $g_{L,R}$ in SL(2,R) can be rewritten in terms of the
Gauss decomposition as follows\footnote{%
For any SL(2,R) group element $g=(a,b;c,d),$ the Gauss decomposition is
given by $X^{+}=b/a,\,\,X^{-}=c/a$. Instead of the exponentials in the
middle factor $\exp $ $[\mp u/(k-2)]$ one could take more generally $%
diag\left( a,a^{-1}\right) ,$ where $a$ can have any sign, unlike the
exponentials. One may carry out the quantization in terms of $a$ instead of $%
u.$ However, in the final analysis the currents depend only on $a^2$ or on $%
S\sim a\partial a^{-1}$ which may be rewritten as $S\sim \left| a\right|
\partial \left| a\right| ^{-1}$ even if $a$ changes sign. Therefore, the
parametrization used here, $\left| a\right| =\exp [-u/(k-2)],$ is adequate
for the general case. }
\begin{equation}
g_{L,R}=\left(
\begin{array}{cc}
1 & 0 \\
X_{L,R}^{-} & 1
\end{array}
\right) \left(
\begin{array}{cc}
:e^{-\frac{u_{L,R}}{k-2}}: & 0 \\
0 & :e^{\frac{u_{L,R}}{k-2}}:
\end{array}
\right) \left(
\begin{array}{cc}
1 & X_{L,R}^{+} \\
0 & 1
\end{array}
\right)   \label{group}
\end{equation}
We compute the left/right currents (omitting the $L,R$ indices for
simplicity)
\begin{equation}
i\left( k-2\right) :\partial _zgg^{-1}:\,=\left(
\begin{array}{cc}
-J^2(z) & J^0(z)+J^1(z) \\
-J^0(z)+J^1(z) & J^2(z)
\end{array}
\right)   \label{cur}
\end{equation}
As compared to the classical currents (\ref{clcur}) we have shifted $%
k\rightarrow \left( k-2\right) \,$ in both $g$ (\ref{group}) and the
definition of the current (\ref{cur}), and applied normal ordering. This
renormalization is necessary for the commutation rules to work out, and is
consistent with similar phenomena concerning the quantization of the WZW
model \cite{ibsfeffaction}. One finds then
\begin{eqnarray}
J^0(z)+J^1(z) &=&\,:\left( k-2\right) \,i\partial _zX^{+}\,e^{-2u/(k-2)}:
\nonumber \\
J^2(z) &=&\left( k-2\right) :i\partial _zX^{+}X^{-}e^{-2u/(k-2)}:+i\partial
_zu  \nonumber \\
J^0(z)-J^1(z) &=&\left( k-2\right) :i\partial _zX^{+}\left( X^{-}\right)
^2e^{-2u/(k-2)}: \\
&&\ \ \ \quad \quad +2X^{-}i\partial _zu-ik\partial _zX^{-}  \nonumber
\end{eqnarray}
The coefficient of $-ik\partial _zX^{-}$ is ambiguous because of the normal
ordering of the term $:i\partial _zX^{+}\left( X^{-}\right)
^2e^{-2u/(k-2)}:\,$. Again this has to be fixed by requiring that the
commutation rules work out. Therefore, instead of having naively $-i\left(
k-2\right) \partial _zX^{-},$ we actually must have $-ik\partial _zX^{-}.$
These results are established by applying the canonical formalism and
identifying these structures with canonical conjugate variables. Velocities
must be replaced by canonical momenta. Note that for left/right movers $%
\partial _z$ can be related to time derivatives $\partial _\tau $ or space
derivatives $\partial _\sigma $. So, at the quantum level we find that we
must identify the canonical pairs ($X^{-},P^{+})$ and $\left( u,S\right) $
as follows
\begin{eqnarray}
P^{+}(z) &=&\left( k-2\right) i\partial _zX^{+}e^{-2u/(k-2)}
\label{canonical} \\
S(z) &=&i\partial _zu  \nonumber
\end{eqnarray}
and then the currents take the form
\begin{eqnarray}
J^0(z)+J^1(z) &=&P^{+}(z)  \nonumber  \label{wzwcurrents} \\
J^2(z) &=&\,:X^{-}P^{+}:+S  \label{wzwcurrent} \\
J^0(z)-J^1(z) &=&\,:X^{-}P^{+}X^{-}:\,+2SX^{-}-ik\partial _zX^{-}  \nonumber
\end{eqnarray}
This is the form used in the previous section without the extra field $%
L^{\prime }(z).$ Thus, as discussed before, only the principal series will
emerge in the WZW model. Using the oscillator form introduced in (\ref
{fields}$)$ we can express $u(z)$ and $X^{+}(z)$ in terms of the basic
oscillators $s_n,\alpha _n^{+}$ by inverting the formulas in (\ref{canonical}%
), thus
\begin{eqnarray}
u(z) &=&u_0-is_0\ln z+i\sum_{n\neq 0}\frac 1ns_nz^{-n}  \label{ux} \\
X^{+}(z) &=&-i\int^zdz^{\prime }\frac{P^{+}(z^{\prime })}{\left( k-2\right) }%
\,\,:\exp \left[ \frac{2u(z^{\prime })}{k-2}\right] :\,.  \nonumber
\end{eqnarray}
Then these structures satisfy the operator products
\begin{equation}
\begin{array}{l}
<u(z)\,S(w)>=\left( \frac i{z-w}+\frac i{2w}\right) \left( \frac k2-1\right)
\\
\left[ J^0(z)-J^1(z)\right] \times X^{+}(w)\rightarrow \frac{-i}{z-w}%
:e^{2u(w)/(k-2)}:
\end{array}
\end{equation}
Thus, $u(z)$ is just the canonical conjugate to $S(z).$ Another property of $%
X^{+}$ that follows from the fundamental operator products is that it is a
singlet under the action of $J_2(z)$%
\begin{equation}
J_2(z)\times X^{+}(w)\rightarrow 0.  \label{j2xp}
\end{equation}
Actually $\partial X^{+}$ is a screening current (see below). Its operator
products with all the currents is either zero or a total derivative.
Therefore, its zero mode commutes with all the currents.

Inserting the expressions in eq.(\ref{ux}) into (\ref{group}) we obtain the
quantum operator version of the group element $g$. The operator products may
now be evaluated. We find the correct quantum products (\ref{quantumrules})
with the above construction in terms of oscillators. That is,
\begin{equation}
\begin{array}{l}
\left[ J_{L,R}^0(w)+J_{L,R}^1(w)\right] \times g_{L,R}(w)\rightarrow \frac{-i%
}{z-w}\left(
\begin{array}{ll}
0 & 0 \\
1 & 0
\end{array}
\right) g_{L,R}(w) \\
J_{L,R}^2(z)\times g_{L,R}(w)\rightarrow \frac{i/2}{z-w}\left(
\begin{array}{cc}
1 & 0 \\
0 & -1
\end{array}
\right) g_{L,R}(w) \\
\left[ J_{L,R}^0(w)-J_{L,R}^1(w)\right] \times g_{L,R}(w)\rightarrow \frac
i{z-w}\left(
\begin{array}{ll}
0 & 1 \\
0 & 0
\end{array}
\right) g_{L,R}(w)
\end{array}
\end{equation}
This result, combined with the current $\times $ current operator products
that we have proven earlier, is convincing evidence that the free field
formalism that we have discussed corresponds to the quantization of the
SL(2,R) WZW model.

\section{Physical states}

\subsection{No ghosts}

Since we have rewritten the WZW theory in terms of free fields, the space of
states consists of the Fock space for the oscillators $\alpha _n^{\pm },s_n$
applied on the base $|p^{+},p^{-},s_0>$ that diagonalizes the zero mode
operators $\alpha _0^{\pm },s_0.$
\begin{equation}
\prod_{n=1}^\infty \left( \alpha _{-n}^{+}\right) ^{a_n}\prod_{m=1}^\infty
\left( \alpha _{-m}^{-}\right) ^{b_m}\prod_{k=1}^\infty \left( s_{-k}\right)
^{c_k}\,|p^{+},p^{-},s_0>
\end{equation}
where the powers $a_n,b_m,c_k$ are positive integers or zero. This is the
space of states that provide a representation basis for the SL(2,R) currents
with only the principal series. The physical states are identified as those
linear combinations that are annihilated by the the total Virasoro
generators
\begin{equation}
L_n|\psi >=0,\qquad n\geq 1.
\end{equation}
In the present case the total Virasoro generators include the following
terms
\begin{equation}
L_n=L_n^{\pm }+L_n^S.
\end{equation}
where
\begin{equation}
\begin{array}{ll}
L_n^{\pm }= & \sum_m:\alpha _{-m}^{-}\alpha _{n+m}^{+}: \\
L_n^S= & \frac 1{k-2}\left( \sum_m\,:s_{-m}s_{n+m}:\,+ins_n+\frac 14\delta
_{n,0}\right)
\end{array}
\end{equation}
Note that the $L_n^{\pm }$ is equivalent to the $c=2$ Virasoro operator in
2D flat spacetime$.$ The full central charge is
\begin{equation}
c=\frac{3k}{k-2}.
\end{equation}
The eigenvalue of the total $L_0$ is
\begin{equation}
L_0=p^{+}p^{-}+\frac 1{k-2}\left[ s_0^2+1/4\right] +\ {\rm integer}=a
\end{equation}
Thus, the theory has been reduced to a 2D lightcone in flat spacetime plus a
Liouville type space-like free field that has positive norm. A small but
important difference as compared to the standard Liouville formalism is that
the linear term in $L_n^S$ is Hermitean in our case, and does not contribute
to $L_0^S.$

The only negative norm states are the ones produced by the time-like
oscillator $\alpha _n^0=\left( \alpha _n^{+}-\alpha _n^{-}\right) /\sqrt{2}.$
However, this is no worse than the usual flat spacetime case. The space of
physical states is defined by
\begin{equation}
\left( L_n-a\delta _{n,0}\right) |\phi >=0
\end{equation}
with $a\leq 1$ fixed. A proof of no ghosts can now be given by following
step by step the same arguments that prove the no ghost theorem in flat
spacetime \cite{noghost}. There is no need to repeat it here. We only recall
that there are no ghosts as long as $a\leq 1$ and $c\leq 26.$

\subsection{Monodromy}

So far we have not taken into account the physical effects of the $\ln z$
cut in the currents. At first sight, the presence of $\ln z$ in the currents
appears to be contrary to the periodicity requirement of closed strings.
However, this is not true. The periodicity requirement arises as a boundary
condition in the process of minimizing the action. In other words only on
mass shell physical string configurations are required to be periodic. In
the presence of the extra zero mode the Hilbert space is larger. One must
require periodicity in the physical on shell sector. The physical sector is
identified as the subspace of states for which the matrix elements of the
currents are periodic.
\begin{equation}
<phys|J^i(ze^{i2\pi n})|phys^{\prime }>=<phys|J^i(z)|phys^{\prime }>.
\label{monodrinv}
\end{equation}
As described below, under this requirement, it turns out that the extra zero
mode must have quantized eigenvalues in the physical sector. Quantum
mechanically it is possible to impose the monodromy condition simultaneously
with the Virasoro constraints since the latter commute with the monodromy
operator as seen below. These monodromy conditions are easily taken care of
in the free field formalism, thus finally giving a complete physical unitary
spectrum of the SL(2,R) WZW\ model.

To implement the monodromy let us first consider its effect on the currents.
{}From the modified currents in (\ref{newold}) we see that under the monodromy
the currents undergo a linear transformation

\begin{eqnarray}
\begin{array}{l}
\left[ J^0+J^1\right] (ze^{i2\pi n})=\left[ J^0+J^1\right] (z) \\
\left[ J^0-J^1\right] (ze^{i2\pi n})=\left[ J^0-J^1\right] (z)+4\pi n\alpha
_0^{-}\,\,J^2(z) \\
\quad \quad \quad \quad \quad \quad +\left( 2\pi n\alpha _0^{-}\right)
^2\left[ J^0+J^1\right] (z) \\
J^2(ze^{i2\pi n})=J^2(z)+2\pi n\alpha _0^{-}\,\left[ J^0+J^1\right] (z)
\end{array}
\label{monodr}
\end{eqnarray}
Note that on the right hand side one finds $J^a,$ not $\tilde{J}^a.$
Therefore we expect that the right hand side can be rewritten as the adjoint
action with a global SL(2,R) transformation. Since the current $J^0(z)+J^1(z)
$ remains unchanged the generator of this transformation must be the zero
mode of this current. Indeed, since $\alpha _0^{-}$ acts like a number, we
can rewrite the monodromy in the form
\begin{equation}
J^i(ze^{i2\pi n})=e^{-2i\pi n\alpha _0^{-}\left( J_0^0+J_0^1\right)
}J^i(z)\,e^{2i\pi n\alpha _0^{-}\left( J_0^0+J_0^1\right) }
\end{equation}
Therefore physical states that satisfy (\ref{monodrinv}) are the subset of
states that are invariant under the monodromy
\begin{equation}
e^{2i\pi n\alpha _0^{-}\left( J_0^0+J_0^1\right) }|phys>=|phys>.
\end{equation}
In the free boson representation this is easy to implement. Using $\left(
J_0^0+J_0^1\right) =\alpha _0^{+}\,$ this condition is applied on the Fock
space of the free bosons in the form
\begin{equation}
e^{2i\pi n\alpha _0^{-}\alpha _0^{+}}\prod_{n,m,k=1}^\infty \left( \alpha
_{-n}^{+}\right) ^{a_n}\left( \alpha _{-m}^{-}\right) ^{b_m}\left(
s_{-k}\right) ^{c_k}\,|p^{+}p^{-}s_0>
\end{equation}
Since $\alpha _0^{-}\alpha _0^{+}$ commutes with all oscillators, it can be
moved to the right and applied on the base. The result is the quantization
condition
\begin{equation}
\begin{array}{l}
e^{2i\pi n\alpha _0^{-}\alpha _0^{+}}|p^{+},p^{-},s_0>=|p^{+},p^{-},s_0> \\
\alpha _0^{-}\alpha _0^{+}=p^{-}p^{+}=-r,\quad r=0,1,2,\cdots
\end{array}
\end{equation}
We must take negative integers because according to the mass shell condition
$p^{-}p^{+}$ is negative. So, the mass shell condition on physical states at
excitation level $l$ takes the form
\begin{equation}
-r+\frac 1{k-2}\left[ s_0^2+1/4\right] +l=a.  \label{shell}
\end{equation}
It is always possible to satisfy this condition with some value of $s_0$
which is quantized in terms of the positive integers $r,l.$ In terms of the
original Casimir $j(j+1)$ this corresponds to a principal series
representation of SL(2,R) with quantized values of $j$ given by $j=-\frac
12+is_0=-\frac 12\pm i\sqrt{\left( k-2\right) \left( r-l+a\right) -1/4}$
where $r$ must be chosen so that the square root is real.

Therefore, it is sufficient to require that $j$ (or $s_0$) has quantized
values
\begin{equation}
j_n=-\frac 12\pm i\sqrt{\left( k-2\right) \left( n+a\right) -1/4}.
\end{equation}
Then the on-shell physical states automatically satisfy the periodicity
condition.

\subsection{Open and Closed strings}

An open string action $S=\int d\tau \int_0^\pi d\sigma \,L(\tau ,\sigma )$
is minimized by allowing free variation of the end points. For the WZW model
for any group $G$ this produces the boundary terms
\begin{equation}
\delta S=\int d\tau \,\left\{
\begin{array}{c}
\left. Tr\left( \left( \delta gg^{-1}\right) \left( \partial _\sigma
gg^{-1}\right) \right) \right| _\pi  \\
-\left. Tr\left( \left( \delta gg^{-1}\right) \left( \partial _\sigma
gg^{-1}\right) \right) \right| _0
\end{array}
\right\}
\end{equation}
In addition to the equations of motion, these terms must also vanish at each
end of the string. That is,
\begin{equation}
\left. \partial _\sigma gg^{-1}\right| _{\sigma =0}=0=\left. \partial
_\sigma gg^{-1}\right| _{\sigma =\pi }.
\end{equation}
At the conformal critical point the equations of motion are satisfied by the
general form $g(\tau ,\sigma )=g_L(\tau +\sigma )\,g_R^{-1}(\tau -\sigma ).$
Then the boundary conditions require that $g_L$ and $g_R$ be related to each
other by the constraint
\begin{equation}
g_L^{-1}\left( \tau \right) \partial _\tau g_L(\tau )+g_R^{-1}\left( \tau
\right) \partial _\tau g_R(\tau )=0.  \label{relation}
\end{equation}
Furthermore, each term in this equation is required to be periodic. As
discussed in the rest of this paper, we impose periodicity on the physical
states. The relation (\ref{relation}) between $g_L(\tau )$ and $g_R(\tau )$
is not easy to solve explicitly. However, we may carry out the quantum
theory in terms of one current $\hat{J}$
\begin{equation}
\hat{J}(z)=g_L^{-1}\left( z\right) \partial _zg_L(z)=-g_R^{-1}\left(
z\right) \partial _zg_R(z).
\end{equation}
This is neither the left moving current $J_L=\partial g_Lg_L^{-1}$ nor the
right moving one $J_R=\partial g_Rg_R^{-1},$ but is related to them by
transformations involving $g_L$ or $g_R.$
\[
J_L=g_L\hat{J}g_L^{-1},\quad J_R=-g_R\hat{J}g_R^{-1}.
\]
The current $\hat{J}$ generates transformations on the right side of $g_L$
and the left side of $g_R^{-1},$ and the meaning of (\ref{relation}) is that
the total current on both $g_L$ and $g_R$ vanishes at the end points. The
canonical commutation rules for this current are identical to the ones we
have already discussed in the rest of the paper. The stress tensor
constructed from it is equal to the stress tensor constructed from either
the left movers or the right movers
\begin{equation}
Tr(\hat{J}^2)=Tr(J_L^2)=Tr(J_R^2).
\end{equation}
The quantum spectrum is obtained from the properties of $\hat{J}$, whose
mathematical structure is the same as either left movers or right movers as
discussed in the previous sections. For an open string we choose to
parametrize the $\hat{J}$ current in terms of free fields. Thus, the quantum
spectrum of the open string in the SL(2,R) curved spacetime becomes
identical to the spectrum discussed above.

For a closed string we have independent left and right moving sectors. The
full group element is $g=g_L(z)g_R^{-1}(\bar{z})$ and there are left and
right moving currents. Therefore we now need two sets of oscillators, the
left movers $\alpha _n^{\pm },s_n,$ and the right movers $\tilde{\alpha}%
_n^{\pm },\tilde{s}_n.$ So, the direct product Hilbert space has a base
labelled by $|p^{-},p^{+},s_0;\tilde{p}^{-},\tilde{p}^{+},\tilde{s}_0>$ with
$p^{-}p^{+}=-r$ and $\tilde{p}^{-}\tilde{p}^{+}=-\tilde{r}$ to insure that
the currents obey the monodromy conditions in the physical sector. We now
need to figure out if these are all independent labels or if they must be
constrained by physical considerations.

For this purpose we recall that a possible modular invariant is the so
called ``diagonal invariant'' that requires the same unitary representation
labelled by the same $j$ for both left and right movers. This may be
understood as being related to the representation of the full group element $%
D^j(g)=D^j(g_L(z))D^j(g_R^{-1}(\bar{z}))$ which requires the same $j$ for
both left and right movers. Therefore, we must demand $s_0=\tilde{s}_0.$

In addition, we examine $g(z,\bar{z})$ in more detail. Keeping the order of
operators, it may be written in the form
\begin{equation}
g=g_L(z)g_R^{-1}(\bar{z})=\left(
\begin{array}{cc}
u & a \\
-b & v
\end{array}
\right)
\end{equation}
with
\begin{eqnarray}
u &=&e^{\frac{-u_L+u_R}{k-2}}-e^{\frac{-u_L}{k-2}}\left(
X_L^{+}-X_R^{+}\right) X_R^{-}\,\,e^{\frac{-u_R}{k-2}}  \nonumber \\
v &=&e^{\frac{u_L-u_R}{k-2}}+e^{\frac{-u_L}{k-2}}X_L^{-}\left(
X_L^{+}-X_R^{+}\right) \,\,e^{\frac{-u_R}{k-2}}  \nonumber \\
a &=&e^{\frac{-u_L}{k-2}}\left( X_L^{+}-X_R^{+}\right) \,\,e^{\frac{-u_R}{k-2%
}}  \label{uvab} \\
b &=&-\left( X_L^{-}-X_R^{-}\right) e^{\frac{-u_L+u_R}{k-2}}  \nonumber \\
&&\quad +e^{\frac{-u_L}{k-2}}X_L^{-}\left( X_L^{+}-X_R^{+}\right)
X_R^{-}\,\,e^{\frac{-u_R}{k-2}}  \nonumber
\end{eqnarray}
We see that $g$ is not periodic under $\sigma \rightarrow \sigma +2\pi n$
since there are logarithms in the expressions for every $X_{L,R}^{\pm }$%
,\thinspace $u_{L,R}.$ However, provided we impose$\,$ $p^{+}=-\tilde{p}^{+}$
on physical states (to cancel the non-periodic behavior in $X_L^{+}-X_R^{+})$%
, we find that we can rewrite this monodromy in the form
\begin{eqnarray}
\begin{array}{l}
g(ze^{i2\pi n},\bar{z}e^{-i2\pi n})=U\tilde{U}g(z,\bar{z})\tilde{U}%
^{-1}U^{-1} \\
U\tilde{U}=e^{-ip^{+}p^{-}2\pi n}e^{-is_0^22\pi n}\,e^{i\tilde{p}^{+}\tilde{p%
}^{-}2\pi n}e^{i\tilde{s}_0^22\pi n}
\end{array}
\end{eqnarray}
where $p^{+},s_0$ are operators which do not commute with $q^{-},u_0$, and
similarly for right movers (note that we have never introduced a canonical
conjugate to $p^{-}$ (or $\tilde{p}^{-})$). To insure that the matrix
elements of the overall $g$ are consistent with monodromy in the physical
sector it is sufficient to impose the conditions
\begin{equation}
2p^{+}p^{-}+2s_0^2-2\tilde{p}^{+}\tilde{p}^{-}-2\tilde{s}_0^2=2m
\end{equation}
where $m$ is an integer. Since we have already seen that $s_0=\tilde{s}_0$
we find that this condition reduces to $r-\tilde{r}=m$, and does not impose
any additional constraints on $r,\tilde{r}.$ Furthermore, for a closed
string we should also have $L_0-\tilde{L}_0=0$ on the physical states.
According to the mass shell condition (\ref{shell}) this requires $r-l=%
\tilde{r}-\tilde{l}.$

So, modular invariant physical closed string states are labelled at the base
as follows
\begin{equation}
|p^{-},\,\,p^{+},s_0\left( n\right) >\times \,|\tilde{p}^{-},-p^{+},s_0%
\left( n\right) >
\end{equation}
where the following restrictions are imposed
\begin{eqnarray}
\tilde{p}^{+} &=&-p^{+},\,\,\,\,\quad \tilde{s}_0=s_0=s_0\left( n\right)
\nonumber \\
s_0\left( n\right)  &\equiv &\left[ \left( k-2\right) \left( n+a\right)
-1/4\right] ^{1/2} \\
n &=&0,1,2,\cdots
\end{eqnarray}
Then the on-shell physical states automatically satisfy the periodicity
conditions.

\section{Chiral Vertex operators}

In string theory there is a vertex operator corresponding to every physical
state. In the usual algebraic approach one would try to construct a chiral
vertex operator $V_{jm}(z)$ corresponding to the ``tachyon'' state $|jm>.$
In our case we have diagonalized $J_0^0+J_0^1\rightarrow p^{+}$ instead of $%
J_0^0\rightarrow m.$ Hence, our vertex operator is labelled as $%
V_{jp^{+}}(z).\,$ In this basis, we need a vertex operator with the
following operator product properties
\begin{equation}
J^a\left( z\right) \,V_{jp^{+}}(w)\sim \frac 1{z-w}\left[
t^aV_{jp^{+}}(w)\right]
\end{equation}
where $t^a$ is a Hermitean generator of SL(2,R) that acts on the $p^{+}$
label in the unitary principal series labelled by $j=-1/2+is.$ This
representation is given by
\begin{eqnarray}
\left( t^0+t^1\right) V_{jp^{+}}(w) &=&p^{+}V_{jp^{+}}(w)  \nonumber \\
t^2V_{jp^{+}}(w) &=&\left[ \frac 12\left\{ p^{+},i\partial _{p^{+}}\right\}
+s\right] V_{jp^{+}}(w) \\
\left( t^0-t^1\right) V_{jp^{+}}(w) &=&\left[ i\partial
_{p^{+}}p^{+}i\partial _{p^{+}}+2si\partial _{p^{+}}\right] V_{jp^{+}}(w)
\nonumber
\end{eqnarray}
A vertex operator with these properties is given by
\begin{equation}
V_{jp^{+}}(w)=\exp \left( ip^{+}X^{-}(w)\right) :\exp \left( \frac{1+2is}{k-2%
}u\left( w\right) \right) :
\end{equation}
It is straightforward to verify that it has the correct operator product
properties with the currents in (\ref{wzwcurrent}). This product of
exponentials has a rather simple structure which can be readily manipulated
in the computation of corellation functions.

There is a $\ln z$ term in the first exponential
\[
\exp \left( p^{+}\alpha _0^{-}\ln w\right) =w^{p^{+}\alpha _0^{-}}
\]
The monodromy condition requires that the power $p^{+}\alpha _0^{-}=\ {%
integer},$ in agreement with the previous approach.

In the construction of corellation functions in other WZW models it has been
found that there are screening charges that play an important role. In the
SL(2,R) case we found the following two screening currents (note that $%
S_1\sim \partial X^{+})$
\begin{eqnarray}
S_1\left( z\right)  &=&\frac{P^{+}\left( z\right) }{k-2}\,:\exp \left( \frac{%
2u\left( z\right) }{k-2}\right) : \\
S_2\left( z\right)  &=&\left( P^{+}\left( z\right) \right) ^{k-2}\,\,:\exp
\left( 2u\left( z\right) \right) :  \nonumber
\end{eqnarray}
Their operator products with the currents are
\begin{eqnarray}
\left[ J^0\left( z\right) +J^1\left( z\right) \right] \,\,S_1\left( w\right)
&\sim &0  \nonumber \\
J^2\left( z\right) \,\,S_1\left( w\right)  &\sim &0 \\
\left[ J^0\left( z\right) -J^1\left( z\right) \right] \,\,S_1\left( w\right)
&\sim &\partial _w\left( \frac{\exp \left[ 2u\left( w\right) /\left(
k-2\right) \right] }{z-w}\right)   \nonumber
\end{eqnarray}
and
\begin{eqnarray}
\left[ J^0\left( z\right) +J^1\left( z\right) \right] \,\,S_2\left( w\right)
&\sim &0  \nonumber \\
J^2\left( z\right) \,\,S_2\left( w\right)  &\sim &0 \\
\left[ J^0\left( z\right) -J^1\left( z\right) \right] \,\,S_2\left( w\right)
&\sim &\partial _w\left( \frac{\left( P^{+}\left( w\right) \right)
^{k-3}e^{2u\left( w\right) }}{z-w}\right)   \nonumber
\end{eqnarray}
Therefore, the zero modes of these screening currents
\begin{equation}
Q_{1,2}\equiv \frac 1{2\pi i}\oint dz\,\,S_{1,2}\left( z\right)
\end{equation}
are the screening charges that commute with all the currents. These play an
important role in the construction of corellation functions. We hope to
discuss these issue further in future work.

\section{Comments}

We have shown that a unitary string theory in SL(2,R) curved spacetime can
be constructed and its spectrum solved exactly. A crucial ingredient was an
additional zero mode whose presence introduces many new technical and
physical features. This should provide a lesson for other more involved
models.

Attention was drawn to some new technical points. The first is that currents
are allowed to contain logarithmic cuts provided monodromy conditions are
applied on the physical states. The second is a new representation of the
currents in terms of free bosons that render the theory completely solvable.
Vertex operators and screening currents that should be useful in
computations were also suggested. These features have obvious
generalizations to higher dimensions as well as to gauged WZW models (coset
models).

We have also shown that the free boson methods permit a more general
representation of SL(2,R) current algebra when the extra degrees of freedom $%
L_n^{\prime }$ are introduced. These were absent in the WZW model, but they
may be present in more general models.

As emphasized in the introduction, the main purpose for the present exercise
is to develop the appropriate methods to study string theory during the
early universe and to understand the impact of string theory on the
symmetries and matter content observed at accelerator energies. For this
purpose the current methods must be generalized to heterotic strings such as
those described in \cite{ibhetero}. Methods used for other special models of
curved spacetimes may also be helpful \cite{others}.



\begin{thebibliography}{99}
\bibitem{BN}  I. Bars and D. Nemeschansky, Nucl. Phys. B348 (1991) 89.

\bibitem{ibcurved}  For a recent review, see: \ I. Bars, ``Curved Space-time
Geometry for Strings ...'', in Perspectives in Mathematical physics''
Vol.III, Eds. R. Penner and S.T.Yau, International Press (1994), page 51.
See also hep-th 930942.

\bibitem{tseytlinrev}  A.A. Tseytlin, reviews, hep-th/9408040, 9410008.
G.T.Horowitz and A.A. Tseytlin, hep-th/9409021. \

\bibitem{O'R}  J. Balog, L. O'Raighfertaigh, P. Forgacs, and A. Wipf, Nucl.
Phys. B325 (1989) 225.

\bibitem{WIT}  E. Witten, Phys. Rev. D44 (1991) 314.

\bibitem{ibsl2wzw}  I. Bars, ``Ghost free spectrum of a quantum string in
SL(2,R) curved spacetime'', hep-th/9503205.

\bibitem{wakim}  M. Wakimoto, Commun. Math. Phys. 104 (1986) 605.

\bibitem{gerasim}  A. Gerasimov, A. Morozov, M. Olshanetsky, A. Marshakov
and S. Shatashvili, Int. J. Mod. Phys. A5 (1990) 2495.

\bibitem{noghost}  \ C. Thorn, Nucl. Phys.\ B248 (1974) 551, and lecture in
Unified String Theory, Eds. D. Gross and M. Green, page 5.

\bibitem{ibstonybrook}  I. Bars, ``Curved Spacetime Strings and Black
Holes'', in Proc. of Strings-91 conference, Strings and Symmetries 1991,
Eds. N. Berkovitz et.al. World Scientific (1992), page 135.

\bibitem{attempts}  P. Petropoulos, Phys. Lett. 236B\ (1990) 151. M.
Henningson, S. Hwang, P. Roberts, and B. Sundborg, Phys. Lett. 267B (1991)
350. J. Distler and P. Nelson, Nucl. Phys. B366 (1991) 255. Shyamoli
Chaudhuri and J.D. Lykken, Nucl.Phys.B396 (1993) 270.

\bibitem{ibsfanomaly}  I. Bars and K. Sfetsos, Mod. Phys. Lett A7 (1992)
1091.

\bibitem{ibjs}  \ I. Bars and J. Schulze, hep-th/9405156, Phys. Rev. D.51
(1995) 1854.

\bibitem{folds}  I. Bars, ``Folded Strings in Curved Spacetime'',
hep-th/9411078, to appear in Phys. Rev D.

\bibitem{halpern}  J.K.Freericks and M.B.Halpern, Ann.of Phys. 188 (1988)
258; Erratum, ibid.190 (1989) 212. M.B. Halpern, E. Kiritsis, N. A. Obers,
K. Clubok, ``Irrational Conformal Field Theory'', hep-th/9501144.

\bibitem{ibsfeffaction}  I. Bars and K. Sfetsos, Phys. Rev. D48 (1993) 844.
A.A. Tseytlin, Nucl. Phys. B399 (1993) 601. A.A. Tseytlin and K. Sfetsos,
Phys. Rev. D49 (1994) 2933.

\bibitem{GSW}  See e.g. Superstring Theory I, M.B. Green, J.S. Schwarz and
E. Witten, pages 111-113.

\bibitem{ibhetero}  I. Bars, Phys. Lett. B293 (1992) 315; Nucl. Phys B334
(1990) 125.

\bibitem{others}  I. Antoniadis, S. Ferrara, S. Kounnas, Nucl. Phys. B421
(1994) 343. E. Kiritsis and C.Kounnas, Phys.Lett.B331 (114) 321. J. Russo
and A. Tseytlin, hep-th/9502038.
\end{thebibliography}
\end{document}